\documentclass[aps,prb,twocolumn,superscriptaddress]{revtex4}
\usepackage{epsfig, graphicx}
\usepackage{bm}

\begin{document}


\title{ Temperature and field dependence of the intrinsic tunnelling structure in overdoped Bi$_{2}$Sr$_{2}$CaCu$_{2}$O$_{8+\delta}$}


\author{T.M. Benseman}

\altaffiliation{Present address: Department of Physics,
Queens College of the City University of New York,
65-30 Kissena Blvd.,
Queens, NY 11367, U.S.A.}
 \author{J.R. Cooper}
\affiliation{Physics Department, Cavendish Laboratory, University
of Cambridge, J.J. Thomson Avenue, CB3 0HE, United Kingdom}

\author{G. Balakrishnan}
\affiliation{Department of Physics, University of Warwick, CV4
7AL, United Kingdom}


\date{\today}

\begin{abstract}
We report intrinsic tunnelling data for mesa structures fabricated
on three over- and optimally-doped
$\rm{Bi_{2.15}Sr_{1.85}CaCu_{2}O_{8+\delta}}$ crystals with
transition temperatures of 86-78~K and 0.16-0.19~holes per CuO$_2$
unit, for a wide range of temperature ($T$) and applied magnetic
field ($H$), primarily focusing on one over-doped crystal(OD80). The
differential conductance above the gap edge shows clear dip
structure which is highly suggestive of strong coupling to a
narrow boson mode.
 Data below the gap edge suggest that
 tunnelling is weaker near the nodes of the $d$-wave gap and give clear evidence for strong
$T$-dependent pair breaking. These findings could help theorists make a
detailed Eliashberg  analysis and thereby contribute towards
understanding the pairing mechanism. We show that for our OD80 crystal the gap above $T_c$ although large, is reasonably consistent with the
theory of superconducting fluctuations.


\end{abstract}


\maketitle

\section{Introduction}
Despite intensive and wide-ranging research in the past thirty
years, detailed understanding of the fundamental physical
properties of high temperature cuprate superconductors, especially
the pairing mechanism, remains elusive. Much of the microscopic
information about their fascinating electronic properties comes
from surface probes such as angle-resolved photo-emission (ARPES)~\cite{Vishik2012}
and scanning tunnelling microscopy (STM)~\cite{Fischer2007,Davis2014,Yazdani2014}, while in the past decade
measurements of transport, e.g. Ref.~\onlinecite{Sebastian2012}  and structural properties, e.g. Refs.~\onlinecite{Forgan2012,Gerber2015} in extremely
high magnetic fields have also been  fruitful. It is important
to verify the results of the surface probes by  bulk measurements
whenever possible. For many years it has been known that mesa
structures fabricated from highly anisotropic
 high-$T_c$ superconductors such as  $\rm{Bi_{2}Sr_{2}CaCu_{2}O_{8+\delta}}$ (Bi-2212)
 may be regarded as stacks of planar ``intrinsic
 tunnel junctions'' (ITJs) connected in series, and  their $I-V$ characteristics
correspond to $c$-axis, superconductor-insulator-superconductor ($SIS$) tunnelling
  spectra \cite{yurgens99}. Planar geometry was used for the ground-breaking tunnelling work on classical superconductors~\cite{Wolf}
and tunnelling in planar ITJs may
  be easier to understand than in
 break junctions \cite{Zasad2011,Mandrus1991,Mandrus1993} where different junctions may sample different regions in $\textbf{k}$-space.  Furthermore one of us has argued~\cite{Cooper2007} that in STM studies the tunnelling probability may have significant  $\textbf{k}$ dependence.   A longer
term goal of the present work is to understand the structure we observe above the gap edge
 and see whether it can be analyzed using Eliashberg theory~\cite{Sui2015,Carbotte2011} to give direct information about a pairing boson.
It will also be important to compare any such results with  Eliashberg analysis of the optical reflectivity~\cite{Carbotte2011} which
 can be performed over a much wider energy range. In the present paper we do not attempt this, but report high quality ITJ data
 and highlight some unexpected findings regarding the  temperature~($T$), voltage~($V$) and magnetic field~($H$), dependence of
 the tunnelling characteristics observed.

In an earlier report~\cite{cond_mat} we showed  experimental data
for ITJs fabricated on two over-doped single crystals of Bi-2212
with $T_c$ values of 80 and 78~K, denoted OD80 and OD78, and an optimally
doped crystal, OP86 with $T_c$=~86~K. Tunnelling results for the
latter crystal and others with  hole concentrations $p<0.19 $
 per CuO$_2$ unit are probably complicated by the presence of the pseudogap, and also of charge density waves that
have been observed  for both Bi-2212~\cite{Yazdani2014, Davis2014}
and YBa$_2$Cu$_3$O$_{6+x}$ (YBCO) with $x$ between 0.45 and 0.93~\cite{Blanco2013}. Here we focus more on OD80,
so our data are complementary to a recent ITJ study \cite{Krasnov2016}
dealing with moderately and slightly underdoped Bi-2212 crystals that do
have a pseudogap.  Our interpretation is  different in that we
suggest that  in OD80 the clear $T$-dependent structure above 2$\Delta_0$, where $\Delta_0(T)$ is the superconducting gap at the anti-nodes,  could
arise from coupling with pairing boson(s) and not from the
pseudogap. For such overdoped crystals ARPES\cite{Vishik2012} and STM\cite{Davis2014} data gives evidence
for a large Fermi surface and no pseudogap at low
$T$, which is in agreement with bulk probes such
as specific heat~\cite{Loram01}, static magnetic susceptibility~\cite{Loram01,Watanabe2000} and
measurements of the London penetration depth~\cite{Tallon2003,Anukool2009}. In
our tunnelling data for OD80 there is evidence for a gap persisting above
$T_c$.  We argue that it is consistent with the microscopic theory~\cite{Larkin} of superconducting fluctuations based on the Ginzburg-Landau free energy expansion, with
relatively small values of the Ginzburg temperature, $\tau_G$.

\section{Methods}

Single crystals of Bi-2212 were grown using a travelling solvent
floating zone furnace and feed rods with nominal stoichiometry of
$\rm{Bi_{2.15}Sr_{1.85}CaCu_{2}O_{8+\delta}}$. These have a
  maximum $T_c$ of $86.5$~K measured by  SQUID magnetometry before fabrication of the mesas and we infer $p$ from the empirical relation~\cite{presland91}
    $T_{c}=T_{c}^{max}(1 - 82.6[p - 0.16]^{2})$, finding $p$ = 0.194, 0.191 and 0.16
    for the three crystals studied. For OD80, $T_c$ measured
    by SQUID magnetometry  agrees to within 1~K with the temperature where $2\Delta_0(T)$, defined by the maxima in  $dI/dV$ curves and shown in Fig.~\ref{fig:2-varyT}(b),
   reaches its minimum value of 34~meV. For OD78 and OP86, the minima in $2\Delta_0(T)$ are 2~K and 6~K lower
    than  $T_c$ values from SQUID magnetometry.  The 6~K discrepancy for OP86 is probably caused by the presence of the pseudogap. This is not a problem  because in Fig.~\ref{fig:logdynes}(a)  the values of $dI/dV$ at high $V$ show that the doping level of mesa OP86 is significantly less than that of OD80, while that of OD78 is slightly larger,  in qualitative agreement with $p-$values obtained from  SQUID magnetometry.\begin{figure}
\includegraphics[width=80mm,height=80mm]{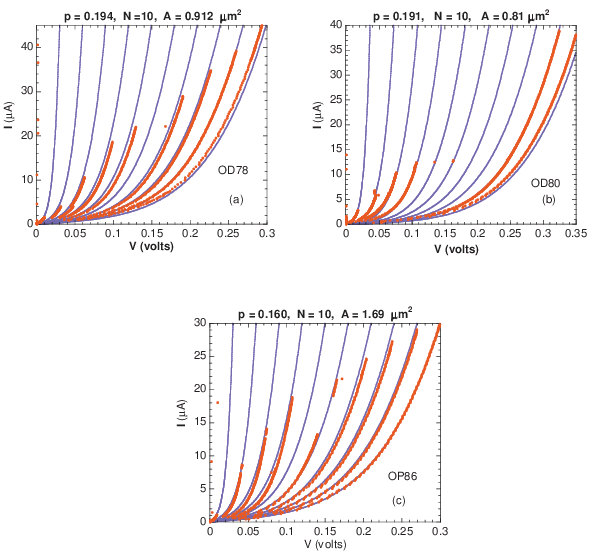}

\caption{Color online: typical $I-V$ curves for the three mesas at low bias
voltages. Red points show data taken at 10~K while increasing $I$ to an appropriate maximum value and then decreasing it. This generates a series of curves in which, from left to right,  the Josephson currents of an increasing number of junctions are suppressed because there is a finite voltage across them.
The blue lines show fits of the form $I = m_1V + m_2V^3 + m_3V^5$
to the (N-1)th curve. The coefficients $m_1$, $m_2$ and $m_3$ are
then scaled by $[(N-1)/n]^i$ where $i$ = 1, 3 and 5 respectively and
$n$ is an integer between 1 and
$N$. Differences between the red data points and the blue lines give
an indication of possible non-uniformity in junction areas, or
more likely, their resistances.}
\label{fig1switching}
\end{figure}
Typical $I-V$ characteristics for
the three mesas at small bias,  taken while sweeping the current
up and down in a controlled manner at 10~K, are shown in
Fig.~\ref{fig1switching}. The branches
      correspond to different numbers
     of Josephson junctions being switched into the resistive state. Switching
to another branch occurs when the critical (Josephson) current of
a particular junction is exceeded and a voltage develops across
it. The computer-controlled current is then swept down to a finite
value before being increased again. When $I$ is large enough, all Josephson
currents are suppressed, there are no further jumps in $V$, and the extreme right hand red curves,
extending to the largest values of $V$ are obtained.  The number of
junctions ($N$) in the stack is equal to
      the total number of branches observed. As shown in Fig.\ref{fig1switching}, these branches scale on to each
      other to a large extent, confirming that the junctions in the mesa have uniform area, and
therefore all junctions switched
      to the resistive state will have the same voltage bias.  However
variations in resistance at the level of 10-15
$\%$ do have significant effects on the magnitude of the structure
in $dI/dV$ above the gap edge. This is a prime cause of a certain
lack of reproducibility in this structure, e.g. between
data  for OD80 and  OD78 in Ref.~\onlinecite{cond_mat} and shown later in  Fig.~\ref{fig:logdynes}(a)
 as well as for under-doped
mesas~\cite{BensemanPhD}.
Mesa $dI/dV$ spectra \cite{BensemanPhD} were measured as the bias
current was swept down from its maximum value towards zero,
thereby maintaining the resistive state. A lock-in technique with
a small 77.7 Hz current modulation was employed, although standard $I-V$ curves were recorded
simultaneously.

 The power dissipation per unit area in HTS mesa structures is
large, sometimes resulting in extreme distortion of $I-V$ curves
by self-heating effects and consequent obliteration of any weak
features in $dI/dV$. Zhu \emph{et al.} \cite{zhu06} have studied
mesa structures in near-optimally doped
 Bi-2212 containing $N=10-11$ junctions in series, finding that there is little
 heating-induced distortion of the $I-V$ characteristic only when the mesa
  area $A$ is $\simeq1\mu m^2$ or less.
Here all three mesas have  $N=10$, the OD78 and OD80 mesas have $A$  below this limit while OP86, although larger, has twice the resistivity
above $T_c$. A high level
of oxygen homogeneity in the mesa is necessary to ensure that any
structure in $dI/dV$ is observed. To avoid possible problems with
ion milling \cite{Krasnov00}, we fabricate our mesas solely by
chemical wet etching \cite{BensemanPhD}.  Finally, irrespective of
the size of the mesas, there is a possibility of electron heating.
For a given $V$ this will  not depend on $N$ or $A$ but only on
the electrical resistance of the junction per unit area and the
thermal resistance for heat transfer between quasi-particles and
phonons. We can rule this out for the OD80 mesa in
Fig.~\ref{fig:3-dip-hump} because the structure  at higher $V$
continues to evolve between 10 and 1.4~K.
\section{Basic Theoretical Background}
The origin of the pairing mechanism in cuprate superconductors
continues to attract the attention of many talented condensed
matter theorists. We hope that some of the points made here will contribute towards their understanding of this problem.   Within the simple ``semiconductor'' picture for
$SIS$  tunnelling~\cite{Wolf} and writing the matrix element for tunnelling
from $\textbf{k}$-space angle $\theta_1$ in electrode 1 to angle
$\theta_2$ in electrode 2 as $M_{\theta_{1}\theta_{2}}$, the
expression for the tunnel current between two identical electrodes
is given by:
\begin{widetext}
 \begin{equation} I\left(V\right)=\int^{2\pi}_{0}\int^{2\pi}_{0}
\int^{\infty}_{-\infty}|M_{\theta_{1}\theta_{2}}|^{\:2}N\left(E,\theta_{1}\right)N\left(E-eV,
\theta_{2}\right)\left[f\left(E-eV\right)-f\left(E\right)\right]dEd\theta_{1}d\theta_{2} \label{eqn:tunintexpl} \end{equation}
\end{widetext}
where $E$ is the energy of a Bogoliubov quasi-particle measured
from the Fermi energy, $N\left(E,\theta\right)$ is the angle-dependent quasi-particle density of states (DOS) whose form in the Dynes approximation~\cite{Dynes}
is given in Eqn.~\ref{eqn:Dynes} and $f$ is the Fermi function. As explained in Ref.~\onlinecite{Eschrig2006} a
distinction needs to be made between incoherent tunnelling where
the  in-plane component of $\textbf{k}$ is not conserved
 and coherent tunnelling where it is approximately conserved, see footnote~\onlinecite{footnote_coh}. For incoherent tunnelling  $|M|^{\:2}$ can
be taken outside the integral and $I$ is given by the product of two angular integrals of the density of states
 factors.  For coherent tunnelling
 $|M|^{\:2}=|M(\theta_1)|^{\:2}\delta(\theta_1-\theta_2)$ and there is only one angular integral.  Theoretically~\cite{Eschrig2006,Andersen}, $M(\theta)^2$
 is expected to vary as
$(\cos{k_x}-\cos{k_y})^4$,   or $(\cos{2\theta})^4$ in
the notation used here.

\begin{figure}
\includegraphics[width=80mm,height=110mm]{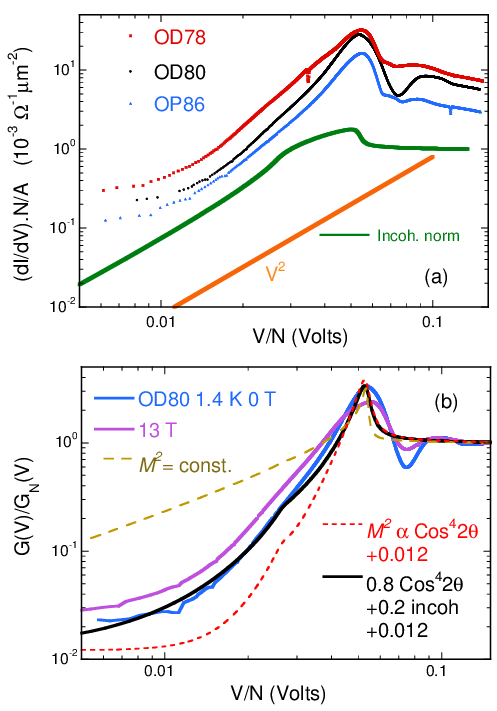}
\caption{Color online: (a) Log-Log plots of $dI/dV$ curves measured for the
three mesas at 1.4 K in zero magnetic field when sweeping $I$ down
from its maximum value. The green curve agrees with an earlier calculation~\cite{Won1994} and shows that purely incoherent
tunnelling gives completely different behaviour. (b) $dI/dV$ for
mesa OD80 at 1.4 K in 0 and 13T fields after normalizing to the
normal state conductance. The solid black curve corresponds to a normalized coherent part calculated from
Eqns.~\ref{eqn:tunintexpl} and~\ref{eqn:Dynes} with $M^2\propto (\cos2\theta)^4$, and multiplied by 0.8 plus a normalized incoherent part multiplied by 0.2. It gives a good description of the data below
$V=\Delta_0$ after adding a small residual term, 0.012, to $G(V)/G_N(V)$. The longer and shorter dashed curves show the calculations for purely coherent tunnelling with
$M^2$ constant and  $M^2\propto (\cos2\theta)^4$ respectively.} \label{fig:logdynes}
\end{figure}

 As shown in Figs.~\ref{fig:logdynes}(a) and (b) the $dI/dV$ curve calculated for incoherent tunnelling has a completely different shape to that for coherent tunnelling and the latter is more similar to our experimental data.
 For completeness, in Fig.~\ref{fig:logdynes}(b) we also show the case where $M(\theta)^2$ is constant to illustrate the contrast with incoherent tunnelling shown in Fig.~\ref{fig:logdynes}(a). It has been
argued~\cite{Eschrig2006} that in the coherent case the anti-nodal
states completely dominate the overall $ G(V)\equiv dI/dV$ curves.
We think this viewpoint
 needs further evaluation because it depends on the presence of a substantial anti-nodal Van Hove singularity deduced~\cite{Eschrig2006}
from ARPES studies,
 which as pointed out by Loram~\cite{LoramPriv} may not be not  consistent with the weak $T$-dependence of the paramagnetic susceptibility~\cite{Loram01,Watanabe2000}. In later discussion, for simplicity,  we consider a cylindrical Fermi surface for which there is no Van Hove singularity.
  Previous work
on ITJs\cite{Latyshev99,Latyshev2000,Vekhter2000} also concluded that there
was a certain amount of coherent tunnelling, but only at the level\cite{Latyshev99}
of 10$\%$. The theoretical curves in Fig.\ref{fig:logdynes} were obtained
using the Dynes formula~\cite{Dynes}
   for $N(E,\theta)$ of a $d$-wave superconductor, namely:

\begin{equation} N(E,\theta) =n(0,\theta) Re[\frac{|E|-i\Gamma}{\sqrt{(|E|-i\Gamma)^2
-(\Delta_0\cos{2\theta})^2}}]\label{eqn:Dynes} \end{equation}Here
$\Gamma$ is the Dynes damping factor and $n(0,\theta)$ is the normal state DOS per unit energy per spin per radian at the Fermi energy. For an isotropic, cylindrical Fermi surface, $n(0,\theta)=n(0)/(2\pi)$, where $n(0)$ is the normal state DOS per unit energy per spin.   The Dynes  formula was  originally used to extract the lifetimes (recombination rates) of excited quasi-particles in classical superconductors~\cite{Dynes} from the $T$-dependent broadening of tunnelling curves.
 It is somewhat different from the formula used to describe various pair-breaking effects  in classical superconductors~\cite{TinkhamBook}, for example by magnetic impurities. Namely  the Dynes formula  gives some zero-energy excitations for any  non-zero value of
 $\Gamma/\Delta$, while the pair-breaking formula only  gives zero-energy excitations (referred to as gapless behaviour) when the scattering rate exceeds a certain threshold value.
The calculated curves in Figs.~\ref{fig:logdynes}(a) and (b) correspond to  an empirical formula, $\Gamma= 0.009 + 0.07x^4/(1+x^2)$,
where $x=E/\Delta_0$. We include the $x^4/(1+x^2)$ factor because
if $\Gamma$ is independent of $E$ then the  curve calculated for
the coherent case shows a strong anomaly at $V=\Delta_0$ that is not
observed experimentally. This arises from the joint
 effect of the peak in the DOS at the anti-node, $\theta=0$,
where $E=\Delta_0$  in electrode 1 and the residual DOS at $E$ = 0, caused  by there being non-zero $\Gamma$ at
 the same angle in
electrode 2. The formula used substantially  reduces the anomaly at $V=\Delta_0$ but still does not account for the width of the peaks at $eV=2\Delta_0$. It corresponds approximately to
expectations for  electron-electron scattering in a $d$-wave
 superconductor where the DOS varies as $E$ for $E\lesssim \Delta_0$, has a weak logarithmic singularity at $\Delta_0$ and
 becomes constant at higher $E$. In this case the usual $E^2$ behavior
 for
 electron-electron scattering with a constant ($E$ independent) DOS changes over to $E^4$ at low
 $E$ where there are two extra  factors of $E$ arising from the linear behavior of the DOS. However electron-electron scattering  is not the only possible cause of an energy-dependent damping factor $\Gamma$: in a $d$-wave superconductor there are  unoccupied quasi-particle states  at arbitrarily low energies, so it could arise from  inelastic scattering  of quasi-particles by the pairing bosons.

 As shown in Fig.~\ref{fig:logdynes}(b), an angular
 independent coherent tunnelling matrix element is ruled out and for the
 damping used we can fit our data with the sum of a dominant (80$\%$) coherent term with
 $M^2 \propto (\cos2\theta)^4$, and a smaller (20$\%$) incoherent contribution.
Eqns.~\ref{eqn:tunintexpl} and~\ref{eqn:Dynes} give a coherent contribution
 $G_S(0)=\int M(\theta)^2[n(0,\theta)\Gamma(\theta)/\Delta(\theta)]^2d \theta$ per spin channel for regions of the Fermi surface
   with $\Gamma(\theta)\lesssim \Delta(\theta)$.
  while quasi-particles in regions
  where $\Gamma (\theta)\gtrsim \Delta(\theta)$ will be essentially  normal. For a $d$-wave superconductor with an angle-dependent gap
       $\Delta(\theta)=\Delta_0\cos2\theta$, such normal regions will have an angular spread of $\pm\Gamma/(2\Delta_0)$ radians around each node and contribute  $\delta n(0)\equiv(2/\pi)\Gamma n(0)/\Delta_0$ to the DOS of an isotropic cylindrical Fermi surface.
       Usually their $\textbf{k}$ states  will be mixed by scattering and their
     tunnelling will be effectively incoherent, giving a contribution of $<M^2>\delta n(0)^2$ to $G_S(0)$.  Because  $M^2\propto(\cos2\theta)^4$, $<M^2>$, its average value  near the nodes,  will be small and  as explained later, any contribution to $G_S(0)$ from quasi-normal regions near the nodes will be dominated by the 20$\%$ incoherent part shown up by the fit in Fig.~\ref{fig:logdynes}(b).

 The log-log plots in
Fig.~\ref{fig:logdynes}(a) show the overall reproducibility of
 $dI/dV\equiv G(V)$ curves  rather directly in that the three
curves are essentially parallel.  A key point in any  analysis is the reproducibility of the values
of $G_S(0)$ when normalized to their values at high $V\simeq0.15
V$ or 5.5$\Delta_0$, both for the 3 ITJs in
Fig.~\ref{fig:logdynes}(a) and for data in the
literature~\cite{Latyshev99,Latyshev2000}.
Because of the re-appearance of small
Josephson currents at low $V$ as $I$ is swept down, we have obtained more precise values
of $G_S(0)/G_N(0)$ from the
  $I(V)$ curves measured at the same time as $dI/dV$.  This was done initially by fitting the data  between  0.06 to 0.018 $V$,
    to $I=m_1(V/\Delta_0) + m_2(V/\Delta_0)^3 +m_3(V/\Delta_0)^5$
 with $\Delta_0 = 0.027V$, but later it was found that straight-line fits to
 $I/V = m_1 +m_2(V/\Delta_0)^2$ showed up unwanted jumps from
 Josephson currents more clearly and gave less scatter in the values of $m_1$. In order to convert
$G_S(0)$ into a residual DOS we also take into account the
$V$-dependence of the conductance $G_N(V)$ in the normal state
using  polynomials  given in footnote~\onlinecite{norm}.
   The
   $H$-dependence of $G_S(0)/G_N(0)$ for the three mesas  obtained from the latter $m_1$ values, i.e. straight line fits to plots of
    $I/V$ vs. $V^2$, is shown later in Fig.~\ref{fig:G0vsH}. It can be seen that all 3 mesas are consistent with  $G_S(0)/G_N(0)$ = 0.012$\pm$0.001 at $H$ =0.
It is interesting to compare this  with the
normalized DOS $0.138/1.2=0.115\pm0.005$ obtained from the low
$T$ specific heat data for Bi-2212 in Ref.~\onlinecite{Junod1994},
where the specific heat coefficient $\gamma = 0.138$
mJ/gm-at./K$^2$,  and the estimated normal state value
$\gamma_n=1.2$ mJ/gm-at./K$^2$ at low $T$ given in
Ref.~\onlinecite{Loram01}. This value is also consistent with
microwave conductivity data on two optimally doped Bi-2212
crystals~\cite{Ozcan2006}, which showed a residual normal fluid fraction of 0.11 and 0.12 for the simpler (Drude) analysis, or alternatively
0.15 and 0.16 for a non-Drude one, as well as with various heat capacity studies of
YBCO~\cite{Junod2001,Riggs2011,Marcenat2015}. However in response to a suggestion from one of the referees we have also fitted
data for the lowest voltage branches of the $I-V$ curves shown in Fig.~\ref{fig1switching} to $I/V$ = $\alpha + \beta V ^2$ and compared the coefficients $\alpha$ and $\beta$  with
$m_1$ and $m_2$ obtained on downward sweeps when all 10 junctions are resistive. Details for the 3 mesas are given as a Table in the supplemental material~\cite{SuppMat}, where it can be seen that $m_1$ and $m_2$ are systematically 20 - 40 $\%$ larger than $\alpha$ and $\beta$. It is not clear at present whether this represents an interesting physical effect or whether it could arise from an unwanted extra conductance path (with a resistance of $\simeq$0.4 M$\Omega$) in parallel with the 10 junctions. In either case it implies that the residual conductances estimated from  our tunnelling data are 20 - 40 $\%$ too high. This does not change our overall conclusions since the discrepancies we discuss later are much larger.  Also  our $T$-dependent data are in good agreement with break junction work, for example Fig.~1 of Ref.~\onlinecite{Mandrus1993}. This rules out  possible effects from a parallel conductance path with a strong $T$-dependence that were suggested by one referee.

In the following we consider 0.16 and 0.11 as upper and lower limits to the residual DOS obtained from heat capacity and microwave studies. Previously we ascribed~\cite{cond_mat} this
residual term to pairs being broken around the nodes. But it is ruled out within the Dynes formulation used here because for a cylindrical Fermi surface with a residual DOS, $\delta n(0)/n(0)$, in the range 0.11 to 0.16, there would have to be broken pairs over an angular range $\pm\alpha$ around each node with $\alpha$ ranging from  5 ($0.11\times45$) to 7.2 ($0.16\times45$) degrees.  With $M^2 \propto(\cos{2 \theta})^4$ there is  a large   attenuation factor
given by $\int_0^\alpha\sin(2\alpha)^4d\alpha/\int_0^{\pi/4}\sin(2\alpha)^4d\alpha$, which ranges  from 0.52 to 3.3 $\times 10^{-4}$ for these values of  $\alpha$. The contribution from incoherent tunnelling between nodes would be larger, ranging from (0.11)$^2\times$0.2 to (0.16)$^2\times$0.2, but still a factor of  5  to 2  smaller that our experimental value of $G_S(0)/G_N(0)$  = 0.012.
The above estimates lead to the conclusion
that the residual conductance, specific heat and unpaired electron states are associated with non-nodal regions.
They must have
 larger values of $\Gamma$, but are not necessarily completely normal, and seeing them in ARPES data might be hampered by the bi-layer splitting. One intriguing possibility is that they are associated  with the ``hot spots'' where
 the antiferromagnetic wave-vector
$\textbf{Q}=(\pi/a,\pi/a)$ spans the Fermi surface. Namely electron
states separated by $\textbf{Q}$ are (a) strongly scattered by
spin fluctuations and (b) must themselves combine in order to give
rise to spin fluctuations at this wave vector, in the same way
that electron states
 separated by a nesting vector combine to
give a charge or spin density wave. We note that the residual
specific heat of YBCO crystals is very
similar~\cite{Junod2001,Riggs2011,Marcenat2015} and is also not
understood. Adding the
residual value of $G_S(0)/G_N(0)$~= 0.012 to the calculated $G_S(V)$ curve in
Fig.~\ref{fig:logdynes}(b) is justified within this picture because at low $V$ the calculated curves are dominated by near-nodal contributions.

However high-quality mesa data taken over 18 years
ago~\cite{Latyshev99} and
  analysed theoretically~\cite{Latyshev99,Latyshev2000,Vekhter2000},  \emph{was} interpreted in terms of pair-breaking at the nodes. In this theory, in the completely coherent limit,
  broken pairs   near the nodes give a
 quasi-particle conductivity at zero bias given by:\begin{equation} \sigma_q = 2(e^2/\hbar)t_\perp^2 N(0)s/(\pi\Delta_0)
 \label{eqn:sigmaq} \end{equation} Here, in the
  notation of Ref.~\onlinecite{Latyshev99}, $t_\perp$ is the $c$-axis tunnelling parameter  at the nodes, $N(0)$ is
  the 2D carrier DOS per spin direction in the normal state and $s$ = 15.2 $\times$ 10$^{-8}$ cm is the interlayer spacing. In  Ref.~\onlinecite{Latyshev99} the  additional incoherent contribution to $\sigma_q$ was found to be negligible for much smaller levels (10$\%$) of coherence so we are justified in neglecting it here.  In contrast to the Dynes formulation, the scattering rate
  does not affect  $\sigma_q$  because of cancellation between
  an increase
 in DOS near the nodes caused by scattering and a decrease in the tunnelling probability associated with the broadening of the
 quasi-particle spectral function $A(\textbf{k},E)$. This tunnelling probability effect is absent in Eqn.~\ref{eqn:tunintexpl}.
 Experimentally $\sigma_q$ is in the range
 1-3~(k$\Omega$cm)$^{-1}$, as indeed it is in our mesas, specifically $\sigma_q$ = 1.6 (k$\Omega$cm)$^{-1}$ for OD80.
 However in contrast to Ref.~\onlinecite{Latyshev99} we believe that
 $t_\perp$ must be angle-dependent, because setting $\sigma_q$ = 1.6~(k$\Omega$cm)$^{-1}$ in Eqn.~\ref{eqn:sigmaq} gives a very low value for $<t_\perp^2>$ = 8.4$\times 10^{-4}$  meV$^2$. This is much smaller than what is expected from  the electrical resistivity and its anisotropy at 300~K, $\rho_{ab}=$ 0.22 m$\Omega$cm and $\rho_{c}=$ 1.8 $\Omega$cm ~\cite{Watanabe2000,Forro1993}.  Work on  anisotropic organic conductors~\cite{Forro1993,Soda1977} suggests that in situations where the in-plane conductivity is described by the usual band theory and the out-of-plane conductivity is via tunnelling,  the formula for resistivity anisotropy~($A$) is the same~\cite{Cooper1994}  to within a factor 2, as that given by standard Boltzmann transport theory, namely $A = <v_\parallel^2>/<v_\perp^2>$, where $v_\parallel$ and $v_\perp$ are the in and out of plane Fermi velocities respectively.  Taking the Fermi surface of Bi-2212 to be  a warped cylinder with tight-binding dispersion in the $c$-direction,
 using $v_\parallel$ = 1.6 $\times 10^7$ cm/sec and $A$ = 8200 gives $<t_\perp^2>$ = 0.28 meV$^2$.
 There is some uncertainty here because our mesa data gives larger values
 of $\rho_{c}=$ 8.1 $\Omega$cm at 300~K, corresponding to $<t_\perp ^2>$ = 0.064 meV$^2$.

The large difference between the value of $<t_\perp>$ given by Eqn.~\ref{eqn:sigmaq} for  $\sigma_q$ = 1.6~(k$\Omega$cm)$^{-1}$ and the value from the resistivity anisotropy at 300~K is consistent with  $M$ being highly anisotropic. But as mentioned already, for a cylindrical Fermi surface with a residual DOS between 0.11 and 0.16, the attenuation from  the  $M^2 \propto (\cos2 \theta)^4$ factor  ranges from 5.2$\times$10$^{-5}$ to  3.3$\times$10$^{-4}$. So even for the larger value $<t_\perp^2>$ = 0.28 meV$^2$, $\sigma_q$  given by Eqn.~\ref{eqn:sigmaq} is still a factor of 58 to 9 too low.
 To summarize, if
 we apply Eqns.~\ref{eqn:tunintexpl} and \ref{eqn:Dynes}  then we would
 conclude that the residual conductivity and DOS mainly arises from low-energy
 states well away from the nodes, a conclusion hinted at in
 $H$-dependent
 specific heat work~\cite{Junod2001}.  Further evidence against significant pair-breaking near the nodes comes from ARPES data for OD80, for example from Fig.~2c of Ref.~\onlinecite{Vishik2012} we estimate that any quasi-normal region is less than $\pm$3 degrees around each node. More calculations may be  needed  regarding broken pairs near the nodes because in Ref.~\onlinecite{Latyshev99} the data were analyzed in terms of strong (resonant)
scattering  and a
large pair breaking parameter $\gamma \sim0.1\Delta_0$.  There is evidence from subsequent microwave studies~\cite{Ozcan2006}
  that weaker, small-angle scattering from out-of-plane defects may be dominant in Bi-2212 crystals.

 We note that the
structure above the gap edge for OD78 and OP86 in Fig.~\ref{fig:logdynes}(a) is smaller  than for OD80. We
suggest that this is  not an intrinsic effect, namely it  arises
from small (10$\%$) variations in the resistance of junctions
within a stack.\begin{figure}
\includegraphics[width=75mm,height=60mm]{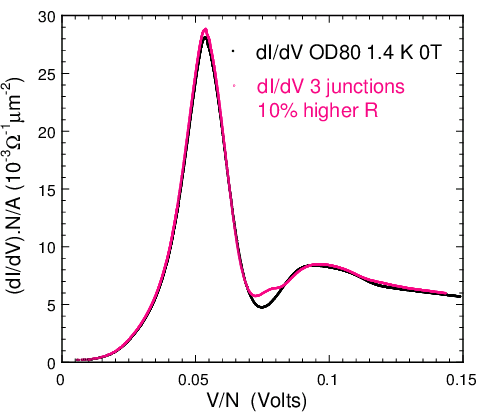}
\caption{Color online: calculation showing sensitivity of structure above the gap
edge to the  resistance of individual junctions. The black curve
corresponds to the case where all 10 junctions have the same
resistance, and $dI/dV$ as measured for the best ITJ OD80. The
purple curve shows the effect of having 3 out of 10 junctions with
10$\%$ higher resistance, i.e. 10$\%$ larger values of $V$ for
the same $I$.} \label{fig:10percent}
\end{figure} As can be seen from Fig.~\ref{fig1switching},
several of the red $I-V$ curves for OP86 and OD78 have  lower
values of $I$ than the blue scaled curves for the same values of
$V$, i.e. their resistances are at least 10$\%$ higher.
Fig.~\ref{fig:10percent} shows that if 3 junctions out of 10 have
10$\%$ higher resistance then this non-uniformity has a strong
effect on the depth and shape of the dip above the gap. We believe
that this is the main reason for a certain lack of reproducibility
in this structure from one ITJ to
another~\cite{BensemanPhD,cond_mat}.

\section{Magnetic field dependence}

\subsection{At zero bias}
 The quasi-particle DOS produced by a magnetic field in
a $d$-wave superconductor at low $T$, is predicted~\cite{Volovik}
to be $N(H)\sim n(0)[H/H_{c2}(0)]^{1/2}$, where $H_{c2}(0)$ is the
upper critical field as $T\rightarrow0$ and $n(0)$ is the
electronic DOS at the Fermi energy in the normal state. This pair
breaking effect arises from Doppler shifts in the  energies of
$+\textbf{k}$ and $-\textbf{k}$ states caused by the superfluid
flow  around the vortices in the vortex state. For low $H$, pairs are broken near the
nodes, where the superconducting gap is small, but the region widens as $H$ is
increased.  The effect is seen in heat capacity studies of
YBa$_2$Cu$_3$O$_7$ crystals, for example
Ref.~\onlinecite{Junod2001}.\begin{figure}
\includegraphics[width=75mm,height=60mm]{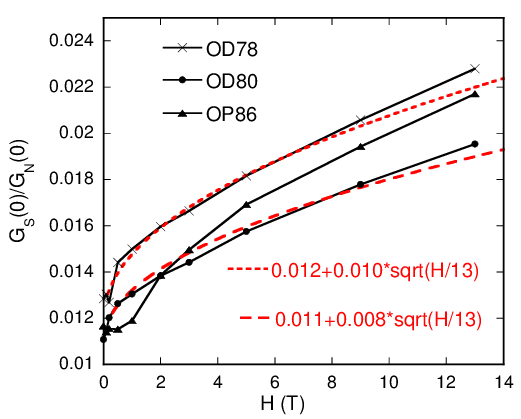}
\caption{Color online: zero bias conductance for the three mesas at 1.4~K obtained from straight line
fits to $I/V$~$vs.$~$V^2$ curves, for $V$ typically  between 0.009 and 0.012 $V$, at various
fields, $H$ applied along the $c$ axis. The normal state
conductance at zero bias $G_N(0)$ is obtained from the
polynomials that give states-conserving fits, see footnote~\onlinecite{norm}. The dashed lines show fits for 20$\%$ incoherent tunnelling of un-paired quasi-particles near the nodes in one layer, generated by the Volovik effect, to zero energy non-nodal states in the neighbouring layer (see text).}
\label{fig:G0vsH}
\end{figure} We estimate $H_{c2}(0)\parallel c$
for OD80 using the clean limit formula
$H_{c2}=0.59\Phi_0/[2\pi\xi_{ab}(0)^2]$ given in Ref.~\onlinecite{Larkin}, where
$\Phi_0$ is the flux quantum for pairs and $\xi_{ab}(0)$, the
in-plane superconducting coherence length as $T\rightarrow0$,
equals $\hbar v_F/[\pi\Delta(0)]$. Here $v_F$ is the Fermi velocity and
$\Delta(0)$ the superconducting gap parameter as $T\rightarrow 0$.
We estimate $v_F$ = 1.58 $\pm0.12 \times 10^7$ cm/sec by applying a
simple cylindrical Fermi surface model to quantum oscillation
data~\cite{Bangura2010} for overdoped Tl2201 crystals for which
the doping-independent effective mass is 5.2$\pm0.4 m_e$. Taking
the measured value $\Delta(0)$ = 26.9~meV at the antinodes  for
our OD80 Bi-2212 mesa, and applying the above formulae, which may
contain extra constants of order unity for $d$-wave rather than
$s$-wave pairing, gives $\xi_{ab}(0)$ = 12.3$\pm0.9 \times 10^{-8}$ cm
and $H_{c2}(0)$ = 128$\pm20$ T.  By analysing specific heat data
for YBa$_2$Cu$_3$O$_7$ crystals the authors of
Ref.~\onlinecite{Junod2001} found $N(H)/n(0) = a\sqrt{8H/[\pi
H_{c2}(0)]}$, where experimentally the constant $a$ = 0.7$\pm23\%$.
According to this formula and the above value of $H_{c2}(0)$ we
would expect the field-induced DOS to be 36$\pm9\%$ of the normal
state value $n(0)$ at 13~T. Plots of $G_S(0)/G_N(0)$ $vs.$ $H$ for the three mesas are shown in
Fig.~\ref{fig:G0vsH}. They all have the same general shape and magnitude.
 both of which agree rather well with previous ITJ experiments~\cite{Latyshev2000}.  However the increase in
$G_S(0)/G_N(0)$ between 0 and 13~T, $\simeq$~0.01, is very small compared with the 36$\pm9\%$ increase in  DOS predicted by the theory of Volovik~\cite{Volovik}. In Refs.~\onlinecite{Latyshev2000} and \onlinecite{Vekhter2000} this was ascribed
 to cancellation between the increased DOS and the increased scattering
 of quasi-particles on vortices, when vortex pancakes in adjacent layers are uncorrelated.

  The above estimates of the constant $a$ and $H_{c2}(0)$ show that for a cylindrical Fermi surface, at 13~T
 pairs should be broken over an angular range of 0.36$\pm0.09\times 45$ = 16.2$\pm4.0$ degrees
 either side of a $d$-wave node. We have considered three simpler interpretations of the $H$-dependence  in Fig.~\ref{fig:G0vsH}, (i) coherent tunnelling, (ii) incoherent tunnelling,  both between  nodal regions in neighboring layers and (iii) incoherent tunnelling between a nodal region in one layer and  non-nodal regions in the neighboring layer. We use the same coherence/incoherence ratio (4:1) as before. Because of the $M^2 \propto(\cos{2 \theta})^4$ factor, case (i) gives a very wide range of values for the increase in $G_S(0)/G_N(0)$ from 0 to 13~T, the upper limit (0.007) is  somewhat lower than  the experimental value, and the $H$ dependence, dominated by  the $M^2 \propto(\cos{2 \theta})^4$ factor, is completely wrong. Case (ii) gives values which are too high by a factor of 2.7$\pm1.3$ and a linear $H$-dependence.  Case (iii) gives very good agreement with experiment both in magnitude and $H$-dependence as shown by the dashed lines in Fig.~\ref{fig:G0vsH}, namely the increase in $G_S(0)/G_N(0)$  between 0 and 13~T is 0.010$\pm0.004$ and is proportional to  $H^{1/2}$ .  The rather  large error arises from  the uncertainty in the constant $a$ in the formula used for the Volovik DOS, and the uncertainty in the residual DOS at $H=0$.   We note that case (iii) implicitly assumes that the vortex pancakes in neighbouring layers are uncorrelated, and this fact would  suppress the nodal-nodal contributions in  cases (i) and  (ii) which might  otherwise be significant.  So, somewhat surprisingly, the Dynes formulation used here plus the assumption that there is 20$\%$  incoherent tunnelling to zero-energy states well away from the nodes, seems to give  a good description of the $H$-dependence of $G_S(0)/G_N(0)$.

\subsection{At higher bias}
\begin{figure}
\includegraphics[width=75mm,height=65mm]{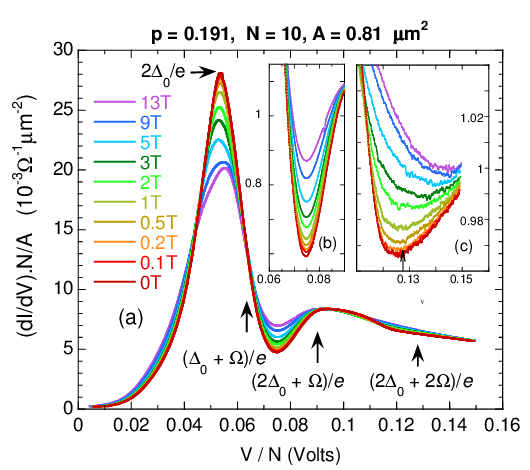}

\caption{Color online: (a) $dI/dV$ curves for OD80 at 1.4~K in
various magnetic fields applied perpendicular to the CuO$_2$
planes plotted $\textit{vs.}$~$V/N$, the bias voltage per
junction. Numerical data is available~\cite{SuppMat}. (b) and (c) show  details of the field dependence of the
lower and upper dips. Here $dI/dV$ curves have been normalized by
dividing through by the polynomial
 given in footnote~\onlinecite{norm}. Values of $(\Delta_0+\Omega)/e$, $(2\Delta_0+ \Omega)/e$
 and 2$(\Delta_0+ \Omega)/e$ are shown by arrows (see text).}
\label{fig:1_2K}
\end{figure}

Fig.~\ref{fig:1_2K}(a) shows $dI/dV$ per unit area for
one junction of OD80, at 1.4 K $\textit{vs}$. the bias voltage per
junction, for many fields $H=$ 0 to 13T applied perpendicular to
the CuO$_2$ planes. The curves are symmetric for $\pm V$, so for
clarity we only show data for $V>0$. For such $SIS$ junctions the
sharp peaks are located at voltages of $2\Delta_{0}/e$, where
$\Delta_{0}$ is the maximum value of the $d$-wave gap.  At 1.4~K
this gives $\Delta_{0}$~=~27.3, 26.9 and 27.4~meV for the three
mesas studied here, OD78, OD80 and OP86 respectively, in good
agreement with the lower values shown in Fig.~15 of
Ref.~\onlinecite{Fischer2007} for these doping levels. The ratio
$2\Delta_{0}/k_BT_c$ = 8.08 $\pm 0.1$, 7.83 $\pm 0.1$ and 7.5$\pm
0.15$ for these three mesas is $\sim$1.75-1.9 times larger than
for a weak-coupling $d$-wave superconductor~\cite{Won1994}.

The $dI/dV$ data for OD80 in Figs.~\ref{fig:1_2K}(a)-(c), show two
$H$-dependent dips above $eV=2\Delta_0$. (Data in
Figs.~\ref{fig:1_2K}(b) and (c) have been normalized see footnote~\onlinecite{norm}).
For a $d$-wave energy gap varying as $\Delta_0\cos2\theta$,
where $\theta $ is the angle between $\textbf{k}$ and the
anti-nodal direction, and for a dispersionless
($\textbf{k}$-independent) boson energy $\Omega$, boson-induced
structure is expected to be most apparent at
$eV$=$2\Delta_0+\Omega$. At this bias voltage, states at the gap
edge at $\Delta_0$ for $\theta = 0$ on one side of the junction
and any
 structure at $\Delta_0+\Omega$ and $\theta = 0$ on the other side are aligned to the same energy,
and strong tunnelling occurs between these. The effect is largest
there because the peak in the superconducting $d$-wave
quasi-particle DOS is largest at $\theta = 0$.  Additional
structure is expected near $eV$=$2\Delta_0+2\Omega$   where
boson-induced anomalies on each side of the junction at $\theta =
0$ have the same energy. However, for a reasonably isotropic Fermi
surface (without a substantial Van Hove singularity as discussed
earlier) we  would expect this structure to extend to lower
energies since at other angles in $\textbf{k}$-space,
$2\Delta_\textbf{k}+\Omega$ will be smaller.

The $S$=1, magnetic resonance excitation, seen by inelastic
neutron scattering~\cite{Keimer2001}, is a candidate pairing boson~\cite{Eschrig2006, Scalapino}. It has an energy $\Omega=5.4k_BT_c$~\cite{Keimer2001,Eschrig2006} and a momentum vector $\bf{Q}$, close to
  ($\pi/a,\pi/a$),~\cite{Keimer2001,Eschrig2006} where $a$ is the in-plane lattice spacing. Various energies
  associated with this value of $\Omega$ are shown in Figs.~\ref{fig:1_2K} and~\ref{fig:3-dip-hump} for OD80.  It can be seen that
   there is a rough correspondence with the
   simple description given above.
In view of the Volovik effect, the
  interpretation of the strong $H$ dependence which we proposed in Ref.~\onlinecite{cond_mat} was  that nodal quasi-particles
  were having a strong effect on  the structure
  above the gap edge. This is still a possibility but  we cannot rule out a much
  more prosaic interpretation in which the disorder associated
  with having uncorrelated vortex pancakes in neighbouring
  layers~\cite{Latyshev2000,Vekhter2000} smooths out this structure.
  Quasiparticles tunnelling from regions between vortices in one layer (where in the first approximation the gap parameter will be the same as
   at $H$=0) to vortex cores in the next layer will give different
   contributions to the $I-V$ curve that depend on the interlayer vortex correlations. If the vortex cores are uncorrelated  this statistical effect will tend to smooth out
   the structure above the gap
 in a similar way to the effect of
  resistance variations shown in Fig.~\ref{fig:10percent}. In
  support of this latter scenario we note that the dip in optimally-doped  Bi-2212
  SIS break junctions~\cite{Mihaly2000} was not suppressed by fields
  of up to 12~T parallel to the $c$-axis. Also recent STM work~\cite{Renner2016} on YBCO shows
  the presence of dips in applied fields of up to 6~T~$\parallel c$ though in the diagrams
  shown they are rather small.

\section{Temperature dependence}

\subsection{Below $T_c$}
\begin{figure}
\includegraphics[width=80mm,height=70mm]{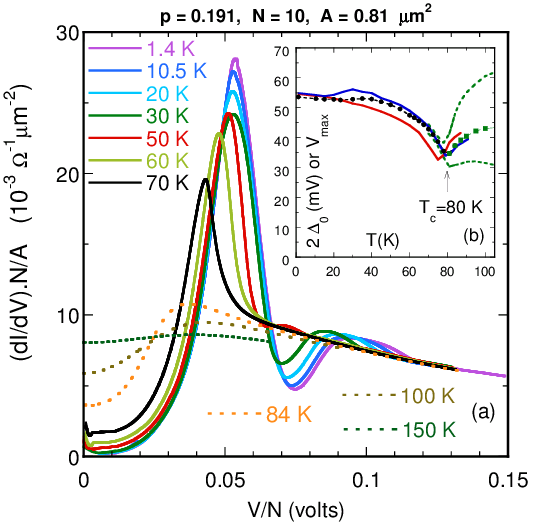}
\caption{Color online: (a) $dI/dV$ curves for OD80 at selected
values of $T$. Numerical data for 31 values of $T$ between 1.4 and 300~K is available~\cite{SuppMat}. (b)
$T$-dependence of the $d$-wave gap 2$\Delta_0(T)$ up to $T_c$ from
the main peaks in $dI/dV$ for OD80 (black circles), OD78 (red
squares) and OP86 (blue triangles). For OD80, green squares above
$T_c$=80~K give voltages of broad maxima in $dI/dV$. Green dashed
lines show their increased breadth by marking regions where
$dI/dV\geq 0.95 (dI/dV)_{MAX}$.}
\label{fig:2-varyT}
\end{figure}

\begin{figure}
\includegraphics[width=80mm,height=65mm]{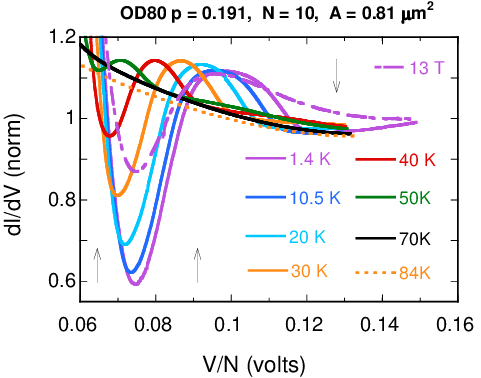}
\caption{Color online:  zoom of the structure above the gap edge,
for  OD80  at selected $T$. The data have
been normalized for clarity, see footnote~\onlinecite{norm}. Values of $(\Delta_0(0)+\Omega)/e$, $(2\Delta_0(0)+ \Omega)/e$
 and 2$(\Delta_0(0)+ \Omega)/e$ are shown by arrows (see text). Data
at 1.4~K for a magnetic field of 13~T $\parallel c$ are also
shown.} \label{fig:3-dip-hump}
\end{figure}

Fig.~\ref{fig:2-varyT}(a) shows the overall $T$-dependence of our raw $dI/dV$ data for OD80 in zero
field at selected temperatures, while data for a total of 31 temperatures below and above $T_c$ are given as supplemental material~\cite{SuppMat}.The data in
Figs.~\ref{fig:2-varyT}(a) and~\ref{fig:3-dip-hump} show  that the dip and the hump at
higher $V$, are strongly $T$-dependent and have almost disappeared
at 50~K even though $\Delta_0$ has hardly changed from its low $T$
value there. The attenuation of the hump is much smaller up to
40~K, but it shifts down with increasing $T$ and also disappears
rapidly between 50 and 60~K. We feel  that this strong
$T$-dependence, especially the shifts of the dips and humps with
$T$, is unlikely to be caused by a conventional phonon pairing
mechanism. As shown in Fig.~\ref{fig:3-dip-hump}, the lower dip is
partially suppressed by a magnetic field, but it is not shifted,
unlike the effect of temperature. For all three mesas a relatively sharp fall in $2\Delta_0(T)$ also sets in just above
   50~K as shown in
      Fig.~\ref{fig:2-varyT}(b). A striking feature of Fig.~\ref{fig:2-varyT}(b) is that
      just below $T_c$,
2$\Delta_0(T) \simeq \Omega=5.4k_BT_c$, possibly suggesting that
the integrity of the magnetic mode is essential for
superconductivity~\cite{Eschrig2006}.

\begin{figure}
\includegraphics[width=80mm,height=120mm]{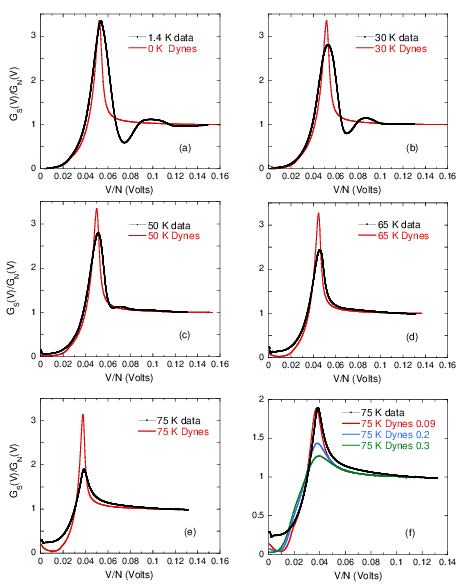}
\caption{Color online:(a) to (e) comparison of normalized $dI/dV$ curves with the Dynes formula Eqn.~\ref{eqn:Dynes}, with a 4:1 coherence-incoherence ratio,
and  the same damping factor,
$\Gamma= 0.009 + 0.07x^4/(1+x^2)$ where $x=E/\Delta_0(T)$, at the selected temperatures shown,  (f) effect of various
extra dampings, $\Gamma= 0.09 + 0.07x^4/(1+x^2)$ etc. at 75~K.  } \label{FigDynesVarT}
\end{figure} The $dI/dV$ curves in Fig.~\ref{fig:2-varyT}(a) \emph{above} $T_c$
also have peaks whose breadth increases rapidly with $T$ as
indicated by the green dashed lines for OD80 in
Fig.~\ref{fig:2-varyT}(b).
 They appear to be states-conserving, for example at 84~K the polynomial
 normalization used earlier
 gives a $dI/dV$ curve that
conserves states to within 2.2$\%$ for the range of $V$ shown in
Fig.~\ref{fig:2-varyT}(a). The presence of these broad peaks for
$p$ = 0.19 agrees with a laser ARPES study~\cite{Vishik2012} of Bi-2212 showing a
  a pseudogap above $T_c$
extending up to $p=0.22$. However in a later section we show that
for OD80 and presumably for the $p$ = 0.19 crystal studied by
ARPES~\cite{Vishik2012}, the pseudogap above $T_c$ is consistent
with the effect expected from  superconducting
fluctuations~\cite{Larkin}. In contrast   the ``real'' pseudogap, which we believe to be an energy scale,
sets in abruptly~\cite{Vishik2012} below $p$ = 0.19 in agreement
with earlier heat capacity~\cite{Loram01} and penetration depth~\cite{Tallon2003,Anukool2009} measurements and is not expected to be visible in
our data for OD80. Furthermore, in contrast to the gap above $T_c$ discussed here, the  ``real'' pseudogap is not states-conserving~\cite{Loram01}.

As recognized previously~\cite{Fischer2007,Mandrus1993}, the $T$-dependence
of the data in Fig.~\ref{fig:2-varyT}(a) at all bias voltages
below 0.12 V  cannot be ascribed simply to thermal broadening.
In Ref.~\onlinecite{cond_mat} we argued that this could be shown in a model-independent way by comparing measured $dI/dV$ curves at a given $T$ with
the 1.4~K curve smoothed over an appropriate voltage range corresponding to $eV$ = 5.6 $k_BT$. However calculations using  Eqn.~\ref{eqn:Dynes}
at various temperatures do not support this procedure, so in Figs.~\ref{FigDynesVarT}(a) to (e) we show instead comparisons of our data with Eqn.~\ref{eqn:Dynes}
at selected values of $T$. The calculated curves all have the same ($E$-dependent) values of $\Gamma= 0.009 + 0.07x^4/(1+x^2)$,
used in Fig.~\ref{fig:logdynes}(b), but now $x=E/\Delta_0(T)$, with  $\Delta_0(T)$ given in Fig.~\ref{fig:2-varyT}(b). In Fig.~\ref{FigDynesVarT}(f)  we show the effect of
 extra $E$-independent damping values, i.e.  $\Gamma= \gamma + 0.07x^4/(1+x^2)$
with $\gamma$ ranging from 0.09 to 0.3. Fig.~\ref{FigDynesVarT} highlights
the fact that the peaks at $2\Delta_0(T)$ become narrower at the same temperature,  near 50~K, where the dip-hump structure is strongly attenuated.
It implies that below 50~K this structure and the broadening have a common origin, namely renormalization from the pairing boson(s).

In Figs.~\ref{FigDynesVarT}(d) and (e) the upturn in the Dynes calculation
near $V$ = 0 is clearly seen,  This
 is well known in classical $SIS$ tunnelling work~\cite{Wolf} and arises from the thermal population of quasi-particle
states above and below the gap edge when $\Delta\sim k_BT$.
But the experimental data show an important difference in that $G(V)$ remains constant  up to larger voltages $\sim0.02V$ than the calculated curves before merging smoothly
with them. This must mean that $\Gamma$ is relatively large where $\Delta$ is large and where $M$ is only weakly
dependent on $\theta$.  We are therefore justified
in estimating a scattering rate $\Gamma$ from the relation $G_S(0,T)/G_N(0)=\Gamma^2/\Delta_0(T)^2$.
\begin{figure}
\includegraphics[width=60mm,height=75mm]{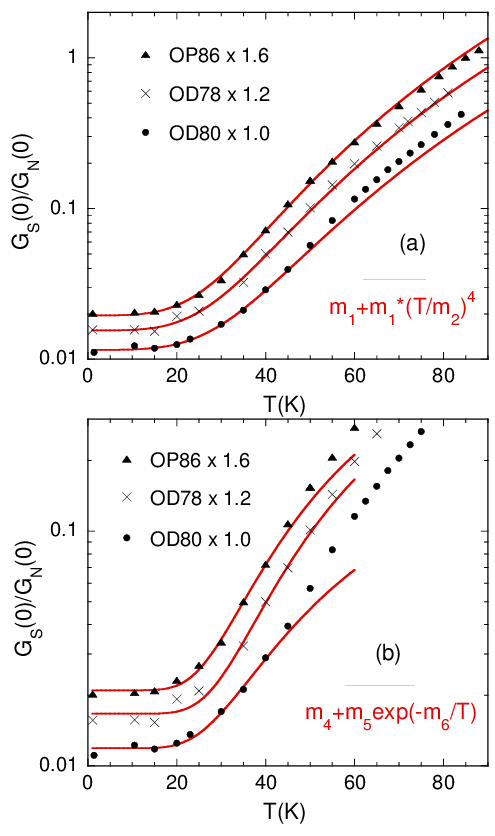}
\caption{Color online: values of $G_S(0)/G_N(0)$ $vs.$ temperature for the 3 mesas. (a) shows the data together with fits to $m_1 + m_1(T/m_3)^4$ with $m_1$=0.0129, 0.0115 and 0.0122 and $m_3$= 33, 36 and 31~K for mesas OD78, OD80 and OP86 respectively. (b) shows the same data and fits to the  Arrhenius law shown with $m_4$ = 0.0139, 0.0119 and 0.0131, $m_5$ = 2.60, 0.65 and 1.67, $m_6$ = 182, 158 and 147~K for mesas OD78, OD80 and OP86 respectively. All fits have been made from 1.4 to 40 K. For clarity the data for OD78 and OP86 have been displaced along  the logarithmic $y$-axis by the multiplying factors shown. } \label{FigGVarT}
\end{figure}
The $T$-dependence of $G_S(0)/G_N(0)$ was obtained by fitting $I-V$ curves at all temperatures measured to $I/V=m_1+m_2V^2$ typically from 0.01 to 0.014 Volts and normalizing  $m_1$ to  the normal state using  states-conserving polynomials~\cite{norm}. The results are shown in Fig.~\ref{FigGVarT} on a semi-logarithmic scale to emphasise the behaviour at low $T$. Normalized  data for the three mesas are in excellent
agreement, the data for OD78 and OD80 obey an $A+BT^4$ law, while
for OP86, $A+BT^3$ gives a marginally better fit. A  $T^3$ power
law was obtained earlier,
 using much larger 20$\times$20$\mu m^2$ mesas, but  employing a pulse method to reduce heating effects and
  suppressing Josephson currents with a 1T applied field~\cite{Suzuki99}.  Fig.~\ref{FigGVarT} is appropriate for two types of comparison with theory. Firstly, as mentioned above, in Eqn.~\ref{eqn:sigmaq}, the scattering rate does not affect $G_S(0)$ and so in this case the main $T$-dependence will presumably come from the fact that normal regions around the nodes expand as $T$ increases causing  $M$ to increase strongly with $T$. However in contrast to the assumption in Ref.~\onlinecite{Latyshev99}, analysis of microwave conductivity data~\cite{Ozcan2006} points towards the importance of small-angle scattering processes so this aspect would need to be addressed. Secondly it has been suggested that thermodynamic fluctuations are extremely important in Bi-2212~\cite{Tallon}.  In this case one might expect the activation energy for 2-dimensional fluctuating normal regions at low $T$  to be given by $E_A = \Delta F_{NS}\xi_{ab}(0)^2s$ where $\Delta F_{NS}$ is the difference in free energy densities at $T=0$ (the superconducting condensation energy density, $U$), $\xi_{ab}(0)$ is the in-plane coherence length at low $T$ and $s$ is the interplanar spacing. Taking $\Delta F_{NS}$ = 1.9~J/gm.at~\cite{Loram01}, $\xi_{ab}(0)$ = 12.8$\times 10^{-8}$ cm and $s = 15.2 \times 10^{-8}$ gives  $E_A/k_B$ = 40~K, of the same order, and actually a factor of 3-4 less than the values obtained from the Arrhenius fits shown in Figs.~\ref{FigGVarT}(b) and \ref{FigsqrtGVarT}(b).   However we argue later that our data for OD80 above $T_c$ are consistent with weaker superconducting fluctuations which goes against this interpretation.

As argued above, within the Dynes formulation used here, the scattering rate $\Gamma$ is given by  $\sqrt{G_S(0)/G_N(0)}\Delta_0(T)$ and appropriate plots are shown in Figs.~\ref{FigsqrtGVarT}(a) and (b).  Fig.~\ref{FigsqrtGVarT}(a) shows the data for the 3 mesas together with fits to an empirical formula describing inelastic scattering between quasi-particles which is expected to vary as $T^4$ at low $T$ and then cross over to $T^2$ as $2k_BT$ becomes comparable with the maximum superconducting energy gap $\Delta_0(T)$. This formula gives a good fit to the data but the crossover temperature $\sim$ 50~K corresponding to 2$k_BT=\Delta_0(T)/3$, could be rather low and  a more precise calculation is needed.  Fig.~\ref{FigsqrtGVarT}(b) shows Arrhenius fits to the scattering rate.
\begin{figure}
\includegraphics[width=60mm,height=75mm]{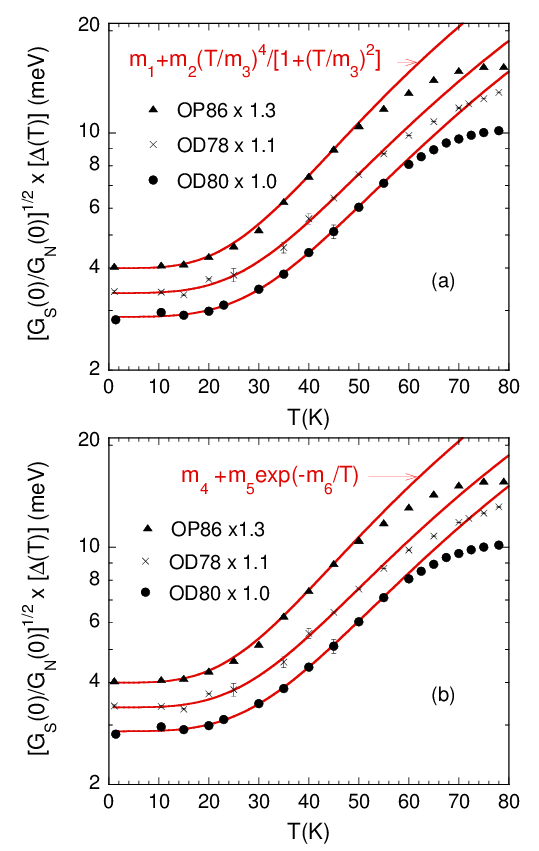}
\caption{Color online: values of $\Gamma(T)\equiv\sqrt{G_S(0)/G_N(0)}\times[\Delta_0(T)]$ $vs.$ temperature for the 3 mesas. (a) shows the data together with fits to the formula shown that is expected for electron-electron scattering  with $m_1$ = 3.07, 2.89 and 3.07~meV, $m_2$ = 7.1, 8.5 and 6.2~meV , $m_3$= 50, 55 and 42~K for mesas OD78, OD80 and OP86 respectively. (b) shows the same data and fits to the  Arrhenius law shown with $m_4$ = 3.13, 2.91 and 3.16 meV, $m_5$ = 56, 49 and 64 meV, $m_6$ =136, 139 and 128 K for mesas OD78, OD80 and OP86 respectively.  All fits have been made from 1.4 to 50~K. For clarity the data for OD78 and OP86 have been displaced along  the logarithmic $y$-axis by the multiplying factors shown.  } \label{FigsqrtGVarT}
\end{figure}  \begin{figure}
\includegraphics[width=70mm,height=55mm]{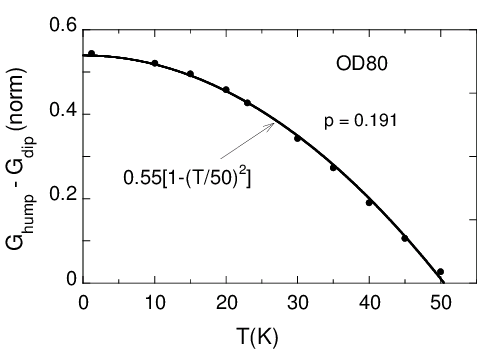}
\caption{ $T$- dependence of the difference between the maximum of $dI/dV$ at the hump and the minimum at the dip for OD80. The points have been obtained from normalized $dI/dV$  curves such as those shown in Fig.~\ref{fig:3-dip-hump}. The data show a clear $T^2$ dependence that goes to zero at 50~K. } \label{Fig:diphumpvsT}
\end{figure}The values of $E_A$ are $\sim$ 12~meV for all three mesas. This is reasonably close to the  energy difference between the $S$ = 1 resonant mode at 5.4$k_BT_c$= 37 meV for OD80 and the value $2\Delta_0$= 54 meV. So another possible interpretation of the activated behaviour is that it represents thermally induced decay of the $S$ = 1 resonant mode  into the quasi-particle continuum above the gap edge.  It could be argued that for a $d$-wave superconductor the gap edge extends down to zero energy at the nodes. But decay processes there would be restricted by $\textbf{k}$ conservation and the fact that the $S$ = 1 mode has a reasonably well-defined $\textbf{Q}$-vector near $(\pi/a,\pi/a)$.
 Fig.~\ref{FigsqrtGVarT}  shows that $\Gamma\sim$~8-10~meV at 60~K for all 3 mesas including OP86. For comparison the scattering rate deduced~\cite{Ozcan2006} from the microwave conductivity of an OP88 Bi-2212 crystal is $\simeq3\times10^{12}$~sec$^{-1}$ at 60~K or  2~meV. From this we conclude that the microwave studies are sensitive to lifetimes nearer the nodes while, because of the strong angle dependence of $M$, our tunnelling data picks up lifetimes nearer the anti-nodes. This could  also be the reason why the $T$-dependence in Fig.~\ref{FigsqrtGVarT} at low $T$ is much flatter than in the microwave studies, e.g. Fig.~4 of Ref.~\onlinecite{Ozcan2006}.
In general, if
real quasi-particle-boson scattering processes are responsible for
$\Gamma$ then its $T$-dependence will be related to the boson
DOS.

We see from Fig.~\ref{FigsqrtGVarT} that the scattering rate for  states near the anti-nodes is $10-12$~meV   as $T_c$ is approached from below, while from Fig.~\ref{fig:2-varyT}(b) $\Delta_0(T\rightarrow T_c)\sim16-17$~meV. We therefore conclude that the larger values of $dI/dV$
at low $V$, i.e. $(dI/dV)_{res}$  are caused by $T$-dependent
pair-breaking processes in line with early break junction
work~\cite{Mandrus1993}. The scattering rates are surprisingly large and must be connected in some way with the fact that as shown in Fig.~\ref{fig:2-varyT}(b) $\Delta_0(T\rightarrow T_c)$ is also large, possibly suggesting that $T_c$ is suppressed by inelastic scattering.
The same general viewpoint was proposed by us in Ref.~\onlinecite{cond_mat} as well as in  two ARPES papers~\cite{Kondo2015,Reber2015} that have inspired a detailed comparison~\cite{Storey2017} of ARPES data with several  bulk properties. But, in contrast to these last three papers, we believe that the gap above $T_c$ in OD80 can be understood reasonably well in terms of the accepted theory~\cite{Larkin} of superconducting fluctuations, as explained below.

 Finally, to conclude this section,  in
Fig.~\ref{Fig:diphumpvsT} we show that the amplitude of the
dip-hump feature for OD80 obeys an $A-BT^2$ law to high accuracy, becoming almost zero at 50~K. Again this seems to be completely at odds with expectations for electron-phonon scattering.
Data for the other two mesas do not show this behavior but in contrast to Ref.~\onlinecite{cond_mat} we now ascribe this to  the  interlayer conductance of OD80 being more uniform than for the other  two mesas.

\subsection{Above $T_c$}

\begin{figure}
\includegraphics[width=80mm,height=120mm]{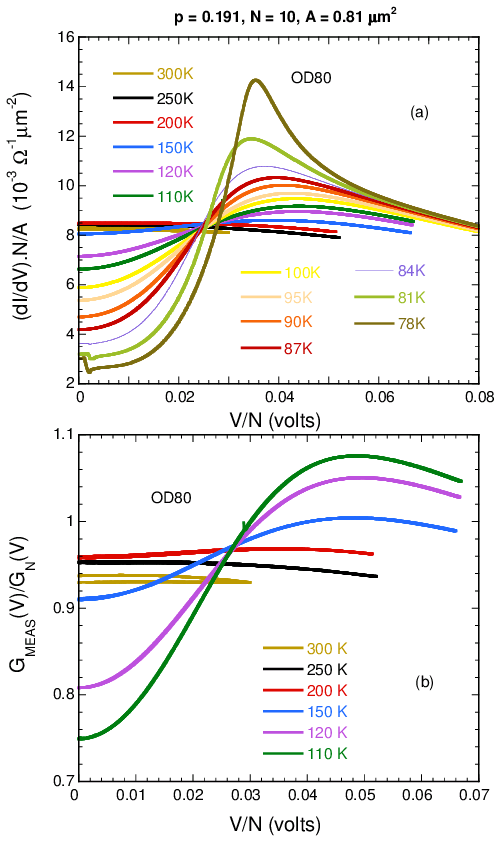}
\caption{Color online:(a) $dI/dV$ curves measured for OD80 above $T_c$ plus one curve at 78~K for comparison. Numerical data for 31 values of $T$ between 1.4 and 300~K is available~\cite{SuppMat}.
(b) Zoom of data measured for OD80 in range 110-300~K, after normalizing by a polynomial that gives a states-conserving curve at 1.4 K and 13~T, see footnote~\onlinecite{norm}. At higher $T$ the sweep voltage range had to be restricted because of hysteresis associated with voltage-induced changes in the mesa resistance~\cite{Muller2011}, as shown for example by the data at 300~K.} \label{fig:OD80highT}
\end{figure}

The $dI/dV$ data for mesa OD80 at all temperatures measured above $T_c$~\cite{SuppMat} are shown in Fig.~\ref{fig:OD80highT}(a) and data at higher $T$, normalized  as explained in footnote~\onlinecite{norm},    are shown in Fig.~\ref{fig:OD80highT}(b).
The theory of superconducting fluctuations~\cite{Larkin} predicts that in the 2D limit the fluctuation contribution to the tunnelling conductance of an $NIS$ junction at $V=0$ is given by:
\begin{equation}G_{FL}(0,T)/G_N(0,T) = -2\tau_G \ln(1/\epsilon) \label{eqn:flucszeroV} \end{equation}
where $G_N(0,T)$ is the conductance in the normal state in the absence of fluctuations, $\tau_G$ is the Ginzburg  parameter and $\epsilon = \ln(T/T_{MF})$, where $T_{MF}$ is the mean field superconducting transition temperature (at which the first term in the Ginzburg-Landau free energy expansion changes sign from positive to negative).
The voltage-dependence of the fluctuation contribution  is given in terms of the second derivative of the digamma function $\Psi$ by~\cite{Larkin}:

\begin{equation}G_{FL}(V,T)\propto G_{FL}(0,T)Re[\Psi^{\prime\prime}(\frac{1}{2}-i\frac{eV}{2\pi k_BT})]\label{eqn:flucsV} \end{equation}
Eqn.~\ref{eqn:flucszeroV}  shows that, unusually, the fluctuation contribution to the conductance has a very weak $T$ dependence because of the double logarithm while Eqn.~\ref{eqn:flucsV} shows that the voltage dependence extends to unexpectedly high voltages since $G_{FL}(V,T)$ has a positive maximum at $eV=\pi k_BT$. In order to make a detailed comparison with our data these equations would need to be extended to the  case of coherent tunnelling between two $d$-wave superconductors with a tunnelling probability $M^2\propto (\cos2\theta)^4$. But to  within  some numerical factors of order unity, 
\begin{figure}
\includegraphics[width=70mm,height=55mm]{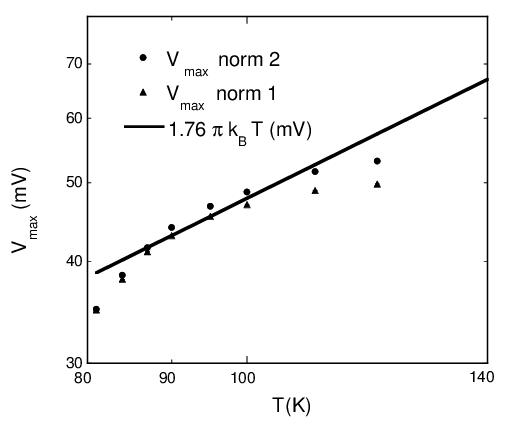}
\caption{Voltages of maxima in normalized $dI/dV$ curves obtained from the data in Fig.~\ref{fig:OD80highT}, by fitting to $a+b(V-V_{max})^2$ near the maxima.   The linear $T$-dependence  expected~\cite{Larkin} from superconducting fluctuation theory has unit slope on the logarithmic scales used. There is indeed a linear region with  approximately the expected slope of $2.0 \pi k_B$ between 87 and 110~K. Results are shown for two different normalizing polynomials~\cite{norm}. } \label{fig:Vmax}
\end{figure}
\begin{figure}
\includegraphics[width=75mm,height=45mm]{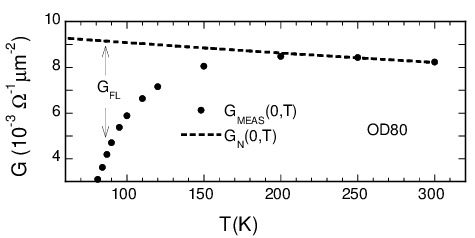}
\caption{Measured values of the zero-bias conductance $vs.$ $T$ at all temperatures above $T_c$. The dashed line shows the estimated normal state conductance background $G_N(0,T)$ obtained by assuming that $1/G_N(0,T) = a +bT$ and finding $a$ and $b$ from the  points at 250 and 300~K. The superconducting fluctuation part, $G_{FL}$ is also shown. } \label{fig:flucdata}
\end{figure}
\begin{figure}
\includegraphics[width=75mm,height=65mm]{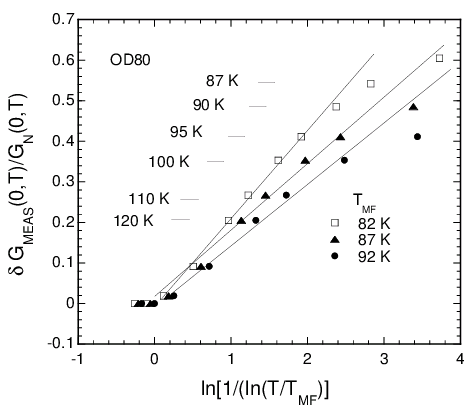}
\caption{Normalized values of the fluctuation contribution to  the zero-bias conductance, $\delta G_{MEAS} = [G_N(V=0,T) -G_{MEAS}(V=0,T)]/G_N(V=0,T)$, obtained from the data in  Fig.~\ref{fig:flucdata}, are plotted $vs.$ $\ln[1/\ln(T/T_{MF})]$ for the three values of $T_{MF}$ shown. The slopes of the lines shown give values of the Ginzburg  parameter $\tau_G$ (see text).
 } \label{fig:fluclog}
\end{figure}
for $SIS$ junctions  we would expect there to be an extra factor of 2 in $G_{FL}(0,T)/G_N(0,T)$~\cite{Larkin} and that the peak in $G_{FL}(V,T)/G_N(V,T)$ will occur at $eV=2\pi k_BT$, a factor of 2 higher than for an $NIS$ junction.   Fig.~\ref{fig:Vmax} shows the voltages  of the maxima in $G_{FL}(V,T)/G_N(V,T)$ $vs.$ $T$ on a log-log plot. The precise positions of the peaks are somewhat dependent on the normal state conductance and we show two limits for this. It can be seen that there is indeed a linear region between 87 and 110~K, and furthermore the slope (1.76$\pi$) is reasonably close to 2$\pi$.

   We have examined the applicability of Eqn.~\ref{eqn:flucszeroV} by subtracting a normal state background contribution of the form $G_N(0,T)=1/(a+bT)$ where $a$ and $b$ are constants fixed by our two measured points at 250 and 300~K, but a quadratic form $G_N(0,T)=a+bT^2$ with $a$ fixed by the first term (8.85)  in the  polynomial used for normalizing $G_S(V,1.4 K)$, and $b$ by the 300~K data point  gave very similar results. The raw data and the $G_N(0,T)=1/(a+bT)$ background are shown in Fig.~\ref{fig:flucdata}. The normalized values $|G_{FL}(0,T)|/G_N(0,T)$ obtained from Fig.~\ref{fig:flucdata} are plotted $vs.$  $\ln[1/\ln(T/T_{MF})]$ in Fig.~\ref{fig:fluclog} for 3 values of $T_{MF}$. Because of the weak dependence of the double logarithm on both $T$ and $T_{MF}$ we cannot use the quality of the straight line fits shown to determine $T_{MF}$.  However, using Eqn.~\ref{eqn:flucszeroV} the slopes of the straight lines shown give values of the 2D Ginzburg temperature ranging from  $\tau_G$ = 0.054 for $T_{MF}$ = 82~K to 0.038 for  $T_{MF}$ = 92~K. The 92~K value is self-consistent in the sense that in the 2D case the suppression of $T_c$ is given by~\cite{Larkin}:\begin{equation} \frac{T_{MF}-T_c}{T_{MF}} = 2\tau_G\ln(1/(4\tau_G) \label{eqn:Tcsupp} \end{equation}
and $\tau_G=0.038$ corresponds to a suppression of 13~K. Because of possible uncertainty in numerical factors in Eqn.~\ref{eqn:flucszeroV} it is worth comparing this estimate of $\tau_G$ with that obtained from the superconducting condensation energy~\cite{Loram01} and the 2D coherence volume mentioned earlier. The original definition of $\tau_G$ was in terms of the reduced temperature above $T_c$ where the Gaussian fluctuation contribution to the heat capacity becomes as large as the mean-field jump that would occur in the absence of fluctuations~\cite{Larkin}. For a classical superconductor with the usual parabolic dependence of the thermodynamic critical field $H_c(T) = H_c(0)[1-(T/T_c)^2]$, the mean field specific heat jump at $T_c$ is $H_c(0)^2/(\pi T_c)$ and setting this equal to the 2D Gaussian fluctuation term $k_B/(4\pi\xi_{ab}(0)^2s\tau_G)$ at $\tau_G$ gives a formula that is probably more general and more suitable for the  $d$-wave superconductor with a larger value of $\Delta(0)/(k_BT_c)$ considered here, namely:
\begin{equation} \tau_G=\frac{k_BT_c}{32\pi U \xi_{ab}(0)^2s }\label{eqn:2Dtau} \end{equation}
where the condensation energy density $U=H_c(0)^2/(8\pi)$.  For OD80 we find  $U \xi_{ab}(0)^2s$  = 40~K, giving a smaller value $\tau_G=0.02$ and from Eqn.~\ref{eqn:Tcsupp}, a suppression in $T_c$ of 10$\%$. Therefore  the line in Fig.~\ref{fig:fluclog} with $T_{MF}$ = 87~K is probably more appropriate than $T_{MF}$ = 92~K.
Recently~\cite{Krasnov2016} the contrasting effects of parallel and perpendicular magnetic fields on underdoped ITJs were used to distinguish the pseudogap from the gap arising from superconducting fluctuations. These authors reached a similar conclusion regarding superconducting  fluctuations but here we have made a more precise numerical comparison with theory~\cite{Larkin} for an overdoped ITJ where  there is no evidence for a pseudogap.

\section{Conclusions}

In summary we have reported  intrinsic $SIS$ planar tunnelling data for three crystals of the cuprate superconductor Bi-2212, and discussed their field- and
temperature-dependence. We believe that there is enough information in our data to assess  the pairing contribution from the $S$ = 1   magnetic mode that
has $\bf{Q}\simeq$~($\pi/a,\pi/a$)~\cite{Eschrig2006,Keimer2001}. On the basis of our analysis using the Dynes equation we conclude that the residual specific heat and normal fluid fraction  do not arise from nodal regions.  We argue that inelastic scattering is large and probably anisotropic since  our tunnelling data,  which are more sensitive to the anti-nodal regions, give scattering rates near 60~K that are approximately 4 times larger than those obtained by microwave studies~\cite{Ozcan2006}. We have discussed the temperature dependence of this scattering  in terms of  electron-electron scattering. However a more exciting possibility is that it is caused  by the same excitations whose virtual exchange is providing
the pairing ``glue''. We have shown that the tunnelling gap above $T_c$  persisting up to 150~K, is reasonably consistent with the theory of superconducting fluctuations~\cite{Larkin} for relatively small values of the 2D reduced Ginzburg temperature, $\tau_G=0.02$.   This is consistent with  the  Gaussian fluctuation analysis used for various cuprates~\cite{Kokanovic2013} that crosses over smoothly to the critical region at approximately $1.02-1.1T_c$.

We would like to thank V. Krasnov  and J. L. Tallon for helpful
discussions and advice, E. J. Tarte and M. Weigand for assistance
with lithographic mask design, staff at the Cambridge Nanosciences
Centre for their help over a long period and A. Carrington and J. L. Tallon  for
comments on the manuscript. The work at the University of Warwick is supported by EPSRC, UK, Grant EP/M028771/1 while that at Cambridge was supported by EPSRC, UK, Grant EP/C511778/1.


\section{Supplemental Material}
\begin{table*}[h]
	\caption{\bf{Supplementary Table} }
		\begin{ruledtabular}
		\begin{tabular}{lllllll}
			\bf{mesa} & $\bm{\alpha \ (\mathrm{10^{-6} A/V})}$ & $\bm{\beta \ (\mathrm{A/V^3}) }$ & \bf{Fit range} \ $\bm{(\mathrm{10^{-4}V^2})}$ & $\bm{m_1}$ \ $\bm{ (\mathrm{10^{-6} A/V}) }$ & $\bm{m_2}$ \ $\bm{(\mathrm{A/V^3})}$ & \bf{Fit range} \ $\bm{(\mathrm{10^{-4}V^{2}})}$ \\
			\hline
			OD80 & $64.7 \pm 1.5$ & $0.29 \pm 0.01$ & $0.5 - 1.9$ & $84.0 \pm 0.3$ & $0.41$ & $0.65 - 2$ \\
			\\
			OD78 & $111 \pm 2$ & $0.48 \pm 0.02$ & $0.2 - 1$ & $160 \pm 2$ & $0.82$ & $0.9 - 1.3$ \\
			\\
			OP86 & $83 \pm 3$ & $0.49 \pm 0.06$ & $0.2 - 0.8$ & $99 \pm 1$ & $0.66$ & $0.8 - 12.4$ \\
		\end{tabular}
		\end{ruledtabular}	
\end{table*}

\nopagebreak
 The ancillary files~\cite{SuppMat} give the numerical data used for making the plots in Figs. 1a,1b, 1c, and Fig. 5 (main). They also contain data for mesa OD80 for 31 temperatures between 1.4 K and 300 K in zero magnetic field some of which are shown in Figs. 6 and 12a.  In the data sets for Figs. 1 the current ($I$) is in units of $\rm{10^{-6} A}$ while in the other data sets it is in $A$. The voltage for the data sets corresponding to Figs. 1 is the measured voltage in Volts while for the other data sets it is the measured voltage divided by 10, i.e. the voltage developed across a single junction. The units of $dI/dV$ are those given in Figs. 5, 6 and 12a.
The table shows the coefficients $\alpha$ and $\beta$ obtained by fitting the lowest voltage branches of the 3 mesas in Fig. 1 to:
\begin{equation} I/V = \alpha + \beta V^2 \label{supeqn:1} \end{equation}
In this case the Josephson current has only been suppressed by an applied voltage across one junction.
A comparison is made with the coefficients $m_1$ and $m_2$ obtained on downward sweeps by fitting data obtained for 10 junctions in series to:
\begin{equation} I/V = m_{1} + m_{2} V^{2} \label{supeqn:2} \end{equation}
In this case the measured voltages have been divided by 10 so that they correspond to the average voltage per junction. This comparison was suggested by one of the referees and is briefly discussed in the m/s.


\begin{references}
\bibitem{norm} We normalize by dividing through  by polynomials $(a + bV^2+cV^4) \times 10^{-3}\Omega^{-1}\mu m^{-2}$. The constants $a$, $b$ and $c$ were  found by requiring
normalized $dI/dV$ curves for  $H$=13~T and 1.4~K, to be unity at
high $V$, 0.142-0.15~V, with zero slope  and to  conserve states between 0 and 0.15~V to within 1$\%$ or better.
 The values of the coefficients are  $a$ = 13.0, 8.85 and 5.305, $b$ = -380.6, -151.27 and -138.75, $c$ =3113, 501.22 and 1670, for mesas OD78, OD80 and OP86 respectively. This procedure was used for the data in Figs.~\ref{fig:logdynes}(b), \ref{fig:1_2K}(b), \ref{fig:1_2K}(c), \ref{fig:3-dip-hump}, \ref{FigDynesVarT}(a) and ~\ref{fig:OD80highT}(b).  The data for OD80 in Figs.~\ref{FigDynesVarT}(b)-(f) are normalized by applying the same procedure to the $dI/dV$ curve for  $H$=0~T and 1.4~K, giving values of $a$ = 9.25, $b$ = -227.6 and $c$ = 3025.6.  In Fig.~\ref{fig:Vmax}, ``norm1'' and ``norm2'' refer to the 13~T and 0~T normalizing polynomials respectively. Choice of the correct background will  be important for  Eliashberg analysis of the dip in $dI/dV$ near 0.13~V shown in Fig.~\ref{fig:1_2K}(c).
\bibitem{SuppMat} Raw data for the 3 mesas in Fig.~\ref{fig1switching} and $I(V)$ and $dI/dV$ data for OD80 in Figs.~\ref{fig:1_2K}, \ref{fig:2-varyT} and \ref{fig:OD80highT}(a) are available as ancillary files.
\end{references}
\end{document}